\documentclass[fleqn,usenatbib]{mnras}

\usepackage{newtxtext,newtxmath}

\usepackage[T1]{fontenc}

\DeclareRobustCommand{\VAN}[3]{#2}
\let\VANthebibliography\thebibliography
\def\thebibliography{\DeclareRobustCommand{\VAN}[3]{##3}\VANthebibliography}

\usepackage{graphicx}	

\usepackage{graphicx}

\usepackage{multirow}
\usepackage{dcolumn}
\newcolumntype{d}[1]{D{.}{.}{#1}}

\usepackage[]{xcolor}

\definecolor{myg}{cmyk}{0.75002,0,1,0}

\definecolor{msnote}{hsb:rgb}{0.492,0.492,0.492}

\newcommand{\hb}{H$\beta$}
\newcommand{\hc}{H$\gamma$}
\newcommand{\feii}{Fe{\sc\,ii}}
\newcommand{\oiiidoublet}{[O{\sc\,iii}]$\lambda\lambda4959,5007$}
\newcommand{\oiii}{[O{\sc\,iii}]$\lambda5007$}
\newcommand{\oii}{[O{\sc\,ii}]$\lambda3727$}
\newcommand{\nev}{[Ne{\sc\,v}]$\lambda3426$}
\newcommand{\neiii}{[Ne{\sc\,iii}]$\lambda3870$}
\newcommand{\heii}{He{\sc\,ii}$\lambda4686$}
\newcommand{\mgii}{Mg{\sc\,ii}}
\newcommand{\mgiidoublet}{Mg{\sc\,ii}$\lambda\lambda2796,2803$}

\newcommand{\kmps}{$\rm km\,s^{-1}$}
\newcommand{\flux}{$\rm erg\,s^{-1}cm^{-2}$}
\newcommand{\lum}{$\rm erg\,s^{-1}$}

\newcommand{\swift}{{\it Swift}}
\newcommand{\fermi}{{\it Fermi}}

\newcommand{\chandra}{{\it Chandra}}

\title[3C~286 classification, variability and SED]{Spectroscopic classification, variability and SED of the \fermi-detected CSS 3C~286: the radio-loudest NLS1 galaxy?}

\author[S. Yao \& S. Komossa]{
Su Yao\thanks{E-mail: syao@mpifr-bonn.mpg.de}
and S. Komossa\thanks{E-mail: astrokomossa@gmx.de}
\\
Max-Planck-Institut f\"ur Radioastronomie, Auf dem H{\"u}gel 69, 53121 Bonn, Germany
}

\date{Accepted XXX. Received 2020; in original form}

\pubyear{2020}

\begin{document}
\label{firstpage}
\pagerange{\pageref{firstpage}--\pageref{lastpage}}
\maketitle

\begin{abstract}
3C~286 is a well-known calibrator source in radio astronomy. 
It is also one of very few compact steep-spectrum sources (CSS) detected in $\gamma$-rays.
Here, we perform a detailed spectroscopic and variability analysis and present the first quasi-simultaneous optical to X-ray spectral energy distribution in order to reveal physical mechanisms which dominate its emission at different wavelengths, and arrive at a reliable optical source classification. 
The first main result of our study reveals several pitfalls when applying simple broad- or narrow-line Seyfert 1 (BLS1 or NLS1) classification criteria which only look at the [O{\sc\,iii}]-H$\beta$ complex. 
[O{\sc\,iii}] and H$\beta$ can be dominated by the same outflow components, in which case FWHM(H$\beta$) is no reliable classification criterion, and extinction by intrinsic or intervening material can make the highest-velocity H$\beta$ component undetectable. 
After careful combination of all information from UV-optical spectra along with multi-wavelength data, we 
confirm
that 3C 286 can be classified as NLS1 galaxy, with line properties and SMBH mass (of order 10$^{8}$ M$_{\odot}$ and accreting near the Eddington limit) close to the BLS1 regime, 
making it an important borderline object. 
The quasi-simultaneous SED taken with \swift\ shows a sharp rise in the optical-UV, 
implying the presence of a strong accretion-disk component with EUV excess, consistent with emission-line diagnostics.
Finally, we report the discovery of X-ray variability of 3C~286, 
plausibly dominated by jet emission, and variable by at least a factor $\sim4$. This result suggests to exercise caution when using 3C~286 as radio calibrator in high-resolution radio VLBI observations. 
\end{abstract}




\section{Introduction}

The majority of sources detected by the \fermi\ mission in  $\gamma$-rays are classical blazars with their jets pointing at us and with massive host galaxies \cite[][]{2020ApJS..247...33A}.
However, \fermi\ has also detected small numbers of other classes of galaxies and active galactic nuclei (AGN), 
including a few starburst galaxies \citep[e.g.,][]{2016ApJ...821L..20P,2016ApJ...823L..17G}, 
nearby Seyfert galaxies \citep[e.g.,][]{2013ApJ...779..131H}, broad-line radio galaxies \citep[e.g.,][]{2011ApJ...740...29K, 2012MNRAS.421.2303B},
and compact-steep-spectral (CSS) sources \citep{2015A&ARv..24....2M,2020ApJS..247...33A}.
In particular, \fermi\ detected several narrow-line type 1 quasars and narrow-line Seyfert 1 galaxies for the first time in $\gamma$-rays
(e.g., \citealt{2009ApJ...707L.142A,
2018ApJ...853L...2P,
2019MNRAS.487L..40Y}; review by \citealt{2018rnls.confE..15K}), 
confirming the existence of relativistic jets in these sources independently inferred from radio studies 
\citep[][]{2006AJ....132..531K,
2008ApJ...685..801Y, 
2015A&A...575A..13F}.

Narrow-line type 1 AGN 
(quasars and their lower-luminosity equivalents of Seyfert galaxies; from now on collectively referred to as NLS1 galaxies) 
are AGN with exceptional properties 
and at an extreme end of a set of correlations between the \hb\ line width and other observables
\citep[][]{1992ApJS...80..109B, 
2000ApJ...536L...5S,
2006ApJS..166..128Z,
2010ApJS..187...64G,
2012AJ....143...83X}.
They are characterized by narrow Balmer lines of the broad-line region (BLR) with full-width-at-half-maximum (FWHM) $<2200$\,\kmps, weak [O{\sc\,iii}] and strong \feii\ emission, 
steep soft X-ray spectra, 
show strong outflow components, 
enhanced star formation activity, 
and are rapidly growing their less massive central supermassive black holes 
\citep[SMBHs; review by][]{2008RMxAC..32...86K}.

Only 7\% of all NLS1 galaxies were found to be radio-loud \citep[preferentially the narrow-line type 1 quasars;][]{2006AJ....132..531K}
and only $\sim$16 of them have been detected in $\gamma$-rays \citep[Table~1 of][]{2018rnls.confE..15K}.
As their multi-wavelength properties are characteristically different from classical blazars, 
their study is important for understanding the disk-jet connection in a previously unexplored parameter regime, and since the known systems are few, it is of great interest to identify more cases. 

The quasar 3C~286 was identified as a CSS  \citep[][]{1982MNRAS.198..843P}
and is exceptional in being one out of only $\sim$5 CSSs significantly detected in $\gamma$-rays with \fermi{\footnote{See evidence for the \fermi\ detection of another CSS-NLS1, RXJ2314.9+2243, but at faint emission levels so its $\gamma$-ray identification still needs to be confirmed
\citep[][]{2015A&A...574A.121K}.}}. 
It is detected at a significance of $\sim6.8\sigma$ 
and a Bayesian-based association probability of 98.5\%
\citep{2020ApJS..247...33A}.
\citet{2020ApJ...899....2Z} re-analyzed the 11-year \fermi\ data, and independently confirmed
the source identification. 

3C~286 is a widely used calibrator source in radio astronomy\footnote{https://casa.nrao.edu/docs/cookbook/}
because of its very stable radio emission and polarization 
for decades
\citep[][]{2013ApJS..204...19P,2013ApJS..206...16P}. 
Its radio emission is compact at sub-arcsecond scales, with a jet and a counter-jet 
\citep[e.g.][]{%
1979ApJ...232..365W, 
1980ApJ...236..707S, 
1997A&A...325..479C, 
2000ApJS..131...95F,
2017MNRAS.466..952A}
and some outer emission extending to $\sim3.8\arcsec$ 
\citep[][]{1989MNRAS.240..657S,1995A&AS..112..235A,2004ChJAA...4...28A}.
The viewing angle of its inner jet on tens of parsec scales was estimated at 48$^{\circ}$ 
based on the apparent proper-motion speed \citep[][]{2017MNRAS.466..952A}.

Because its \hb\ emission was not easily accessible from ground-based spectroscopy given its redshift of $z=0.85$, its 
AGN type remained uncertain, and both an intermediate-type broad-line Seyfert 1 (BLS1) \citep[][]{2010A&A...518A..10V} or a NLS1 classification 
\citep{2016PhDT.yao,2017FrASS...4....8B, 2020MNRAS.491...92L} was suggested. 
Here, we critically investigate the optical spectral classification of 3C~286 based on its full optical-UV emission-line spectrum 
along with tight muti-wavelength constraints on intrinsic absorption, 
evaluate the X-ray variability of special interest with regard to its radio-calibrator status, and present the first simultaneous optical--UV--X-ray spectral energy distribution (SED) of 3C~286. 

We use a cosmology \citep[][]{2006PASP..118.1711W} with 
$H_{\rm 0}$=70 km\,s$^{-1}$\,Mpc$^{-1}$, $\Omega_{\rm M}$=0.3 and $\Omega_{\rm \Lambda}$=0.7 throughout this paper.

\section{Data analysis and results}

\begin{table}
\footnotesize
    \caption[]{Log of SDSS-BOSS, \chandra\ ACIS-S and \swift\ XRT and UVOT observations of 3C 286. 
    }
    \label{tab:obs-log}
    \centering                          
    \begin{tabular}{l c l l r}        
    \hline\hline                 
    \multicolumn{1}{c}{Mission} & 
    \multicolumn{1}{c}{Band} & 
    \multicolumn{1}{c}{Obs-date} & 
    \multicolumn{1}{c}{MJD} &
    \multicolumn{1}{c}{Exp. (s)} \\
    \hline                        
    SDSS  &  &  2005-04-07 & 53467 & 2220 \\ 
    \noalign{\smallskip}
    SDSS-BOSS  & & 2013-02-14 & 56337 & 3603 \\
    \noalign{\smallskip}
    \chandra\ ACIS-S & 0.3-10 keV & 2013-02-26 & 56349  &  2004 \\
    \noalign{\smallskip}
    \swift\ XRT & 0.3-10 keV & 2020-08-16 & 59077 &  2018 \\
                      &                   & 2020-08-21 & 59082 &  1374 \\
    \noalign{\smallskip}
    \swift\ UVOT  
              & {\it v}  & 2020-08-16 & 59077  & 162 \\
              &          & 2020-08-21 & 59082 & 104 \\
              & {\it b}  & 2020-08-16 & 59077 & 162 \\
              &          & 2020-08-21 & 59082 & 104 \\
              & {\it u}  & 2020-08-16 & 59077 & 162 \\
              &          & 2020-08-21 & 59082 & 104 \\
              & {\it w1} & 2020-08-16 & 59077 & 323 \\
              &          & 2020-08-21 & 59082 & 208 \\
              & {\it m2} & 2020-08-16 & 59077 & 485 \\
              &          & 2020-08-21 & 59082 & 373 \\
              & {\it w2} & 2020-08-16 & 59077 & 648 \\
              &          & 2020-08-21 & 59082 & 416 \\
    \hline                                   
\end{tabular}
\end{table}

\subsection{SDSS}\label{sec:sdss_spec}

In the course of the Sloan Digital Sky Survey 
\citep[SDSS;][]{2000AJ....120.1579Y} 
3C~286 was observed on April 7, 2005 (MJD 53467) and February 14, 2013 (MJD 56337), respectively.  
The later spectrum from the SDSS-BOSS survey \citep[][]{2013AJ....145...10D} covers a wider spectral range of $\sim$3600--10400\,\AA\ which corresponds to $\sim$1950--5600\,\AA\ at rest frame of the object.
We retrieved the calibrated 1D spectra of both observations from the SDSS archive 
and corrected them for Galactic extinction with $E(B-V)=0.011$\,mag \citep[][]{2011ApJ...737..103S} and an $R_{V}=3.1$ extinction law. 
Then the spectra were transformed into the source rest frame at $z=0.850$ provided by the SDSS spectroscopic pipeline. 

\begin{figure*}
  \centering
  \includegraphics[width=1.0\textwidth]{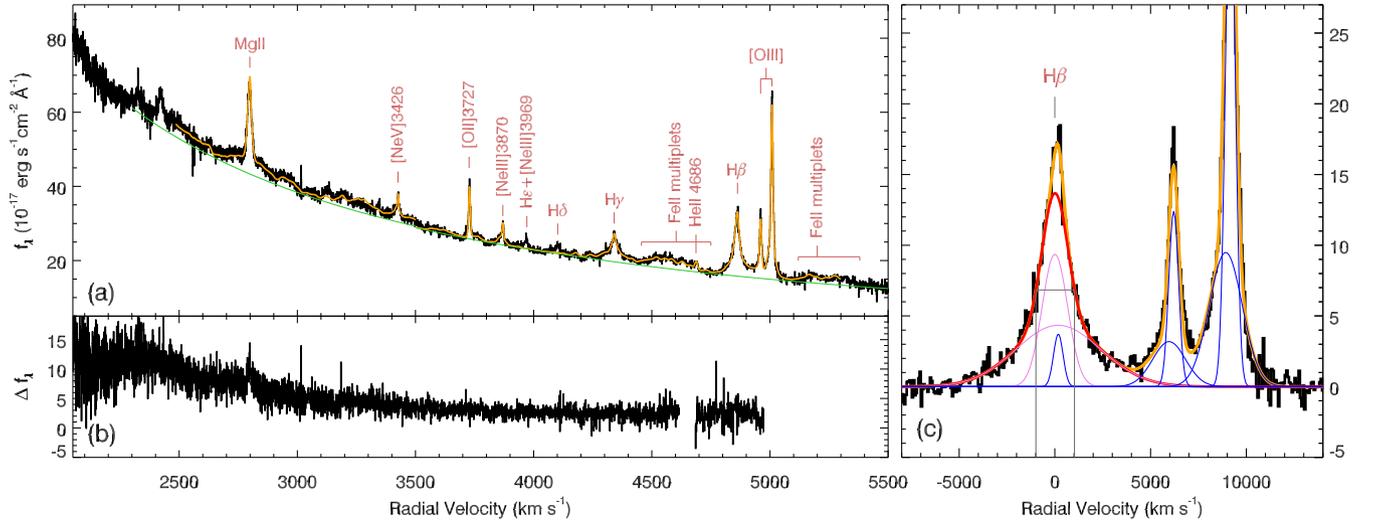}
  \caption{
  (a) The rest-frame SDSS-BOSS spectrum (black) of 3C~286 after correction for Galactic extinction. 
  The orange curves represent the combination of the best-fit models described in Section~\ref{sec:sdss_spec}.
  (b) The residuals of subtracting the 2005 SDSS spectrum from the 2013 SDSS-BOSS spectrum.
  (c) The best-fit decomposition of the \hb-[O{\sc\,iii}] complex. 
  The black curve represents the 2013 SDSS-BOSS spectrum subtracted by the pseudo-continuum model. 
  The red line represents the broad component of \hb\ of which the FWHM is shown by a gray rectangle, 
  and the blue lines represent the narrow component of \hb\ and \oiiidoublet.
  The orange line represents the sum of the models.
  The pink curves represent the double-Gaussian profiles composing the broad \hb. 
  }
  \label{fig:sdss_spec}
\end{figure*}

To inspect the variations between the spectra at the two epochs, 
we scale the spectrum taken in 2005 to match the \oii\ flux of the 2013 spectrum, obtained from direct integration of the continuum-subtracted line profile.
So we assume that the flux of \oii\ does not vary
on timescales of years. 
Then the scaled 2005 spectrum is subtracted from the 2013 spectrum to obtain the spectral variation. 
The spectrum taken in 2013 is shown in Figure~\ref{fig:sdss_spec}(a) and the spectral variation 
is shown in Figure~\ref{fig:sdss_spec}(b). 
As can be seen, 
there is only a minor change in the continuum 
while the emission lines of 3C~286 at the two epochs are nearly constant.
Thus, we only perform a spectral fitting on the 2013 spectrum.

The spectral fitting is based on {\tt IDL} routines in the {\tt MPFIT} package \citep[][]{2009ASPC..411..251M}, 
which performs a $\chi^{2}$ minimization using the Levenberg-Marquardt method. 
Firstly, we fit the [O{\sc\,iii}]-H$\beta$-H$\gamma$ region using data in windows of 
[4720, 5300]\,\AA, 
[4160, 4630]\,\AA, 
[3990, 4080]\,\AA, 
[3900, 3950]\,\AA, 
[3760, 3850]\,\AA, 
[3500, 3690]\,\AA. 
These windows are chosen in order to mask all the possible emission line regions except 
H$\beta$4861, 
H$\gamma$4340, 
[O{\sc\,iii}]$\lambda\lambda$4959,5007 
and the Fe{\sc\,ii} multiplet underneath.

A pseudo continuum consisting of a single power law and \feii\ multiplets modelled by templates of \citet{2004A&A...417..515V} is adopted. 
Each of the \oiiidoublet\ doublet is modelled with two Gaussians, one for the core, the other for the possible blue-shifted wings. 
Both of the doublets are constrained to have the same profile and redshift during the fitting, 
and the flux ratio of [O{\sc\,iii}]$\lambda4959$ to [O{\sc\,iii}]$\lambda5007$ is fixed to the theoretical value of 1:2.98.
The \hb\ and \hc\ emission lines are modelled with three Gaussians. 
One of the Gaussians represents the narrow component originating from the narrow-line region (NLR), and is constrained to have the same profile and redshift as the core of [O{\sc\,iii}]$\lambda5007$. 
The combination of the other two Gaussians represents the broad component presumably originating from the BLR. 
The flux ratio of the narrow \hc\ to \hb\ is fixed at 0.46:1 assuming Case B \citep[][]{2006agna.book.....O}. 
The broad \hb\ and \hc\ lines are constrained to have the same profile and redshift, while their fluxes are set to be free parameters. 
We do not see any visible signature of [O{\sc\,iii}]$\lambda4363$, so this line is not included in the model. 
The best-fit models of \hb\ and \oiiidoublet\ are displayed in Figure~\ref{fig:sdss_spec}(c). 
We also try to use a Lorentzian profile instead of a double-Guassian to model the broad component of \hb, 
which gives an equally good fit result. 

The \mgii\ region is fitted separately from the [O{\sc\,iii}]-H$\beta$-H$\gamma$ region. 
We use the data in the windows of 
[2200, 2295]\,\AA, 
[2355, 2400]\,\AA, 
[2485, 2640]\,\AA, 
[2690, 3100]\,\AA, 
[3170, 3400]\,\AA, 
[3610, 3700]\,\AA\ 
in order to mask the emission line regions listed in Table~2 of \citet{2001AJ....122..549V} except for the \mgiidoublet\ doublet and the \feii\ multiplet underneath. 
Similarly, a single power law plus \feii\ multiplet modelled by the template in \citet{2006ApJ...650...57T} are used to represent a pseudo-continuum. 
Each line of the doublet is modelled with one broad and one narrow component. 
The broad component is a truncated five-parameter Gauss-Hermite profile \citep[][]{2009ApJ...707.1334W} 
and the narrow component is a single Gaussian. 
Both broad components of the doublet are assumed to have the same profile and redshift, 
and their flux ratio is fixed at 1.2:1 assuming an entirely thermalized gas for simplicity
\citep[][]{1997ApJ...489..656L}. 
The narrow components of the doublet are constrained following the same prescription, additionally with their $\rm FWHM\leqslant900$\,\kmps.

\begin{figure}
  \centering
  \includegraphics[width=0.8\columnwidth]{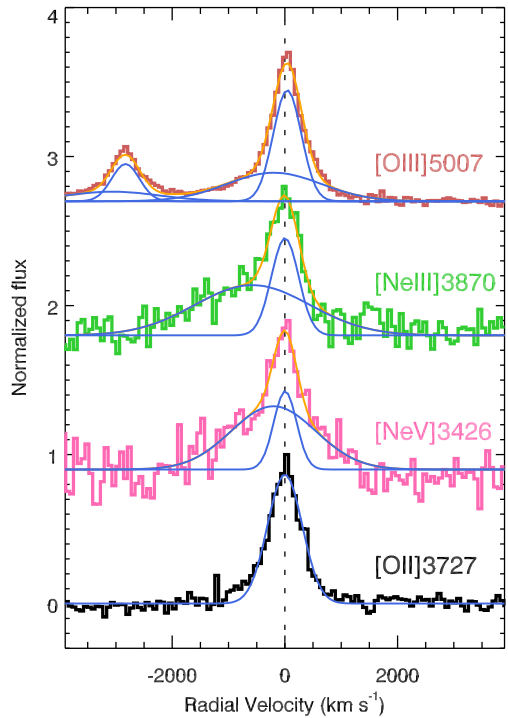}
  \caption{
  The velocity profiles of \oiiidoublet, \neiii\ and \nev\ with respect to \oii. 
  The blue curves represent the best-fit decomposition. 
  The  dotted vertical line denotes the velocity centroid of \oii. 
  A blue wing is present in the high-ionization lines.
              }
              \label{fig:fblines}
\end{figure}

Then we subtract the best-fit model of the [O{\sc\,iii]}-\hb-\hc\ and \mgii\ region from the spectrum, 
and fit \nev, \oii, \neiii\ and \heii\ in the residuals, 
as these lines are visibly detected and isolated, and easy to be measured. 
The \oii\ and \heii\ are modelled by single Gaussian profiles, 
while \nev\ and \neiii\ are decomposed into two Gaussian profiles similarly as \oiii\ considering possible blue-shifted wings. 
The fit results are shown in Figure~\ref{fig:fblines}.
\oii\ is well fitted by a single Gaussian which has a slightly larger line width compared to the core component of \oiii\ (Table~\ref{tab:lines}).
The core components of \nev, \neiii\ and \oiii\ have nearly the same redshifts as \oii.
We also notice very faint excess emission implying that there might be a slight blue wing component in \oii\ which, however, is not statistically significant in the present data.

To estimate the uncertainties of the fitting, 
we follow the same approach as in \citet{2011ApJS..194...45S}. 
We generate 1000 mock spectra by adding Gaussian noise
to the real spectrum flux densities using flux density errors, 
and fit the mock spectra with the same fitting procedure described above.
The uncertainties are determined from the 68\% range of the distributions of fitting results from mock spectra. 
All the measured quantities of the real emission lines and their uncertainties are listed in Table~\ref{tab:lines}.

\begin{table}
    \setlength{\tabcolsep}{6pt} 
    \caption[]{Emission-line results. 
    }
    \label{tab:lines}
    \begin{tabular}{%
                l
                D{,}{\pm}{-1} 
                D{,}{\pm}{-1} 
                D{,}{\pm}{4.4}
                cccccc}
    \hline\hline
        \noalign{\smallskip}
        Line &   
                \multicolumn{1}{c}{FWHM} & 
                \multicolumn{1}{c}{Flux}  &
                \multicolumn{1}{c}{Velocity}  \\
        (1) &   
                \multicolumn{1}{c}{(2)} & 
                \multicolumn{1}{c}{(3)}  &
                \multicolumn{1}{c}{(4)}  \\
        \noalign{\smallskip}
        \hline
        \noalign{\smallskip}
        H$\beta_{\rm broad}$ & 2001, 222 & 119.3, 5.8  &  \cdots  \\
        H$\beta_{\rm broad,lorentzian}$ & 1858, 124 & 133.6, 2.7  &  93, 19  \\
        H$\beta_{\rm broad,comp1}$ & 1556, 246 & 46.8, 8.3  & 136, 38 \\
        H$\beta_{\rm broad,comp2}$ & 5209, 1052 & 72.5, 4.9  & -38, 267\\
        H$\beta_{\rm narrow}$ & \cdots^{a} & 6.9, 2.6  &  \cdots^{a} \\
        H$\gamma_{\rm broad}$ & \cdots^{b} & 42.6, 2.6 & \cdots^{b} \\
        $[$O\,\sc{iii}$]\lambda5007$  & 683, 20 & 133.2, 1.8 &  \cdots \\
        $[$O\,\sc{iii}$]\lambda5007_{\rm core}$  & 572, 29 & 71.5, 5.1  &  -43, 11  \\
        $[$O\,\sc{iii}$]\lambda5007_{\rm wing}$  & 1953, 151 & 61.7, 4.4  & 192.3, 69  \\
        He\,\sc{ii}$\lambda4686$ & 842, 295 & 5.8, 1.0  &  -63, 93  \\
        $[$Ne\,\sc{iii}$]\lambda3870$ & 746, 90 & 18.8, 1.4  &  \cdots \\
        $[$Ne\,\sc{iii}$]\lambda3870_{\rm core}$ & 556, 76 & 5.9, 1.1  & -3, 32 \\
        $[$Ne\,\sc{iii}$]\lambda3870_{\rm wing}$ & 2398, 407 & 12.9, 1.9  &  579, 156 \\
        $[$O\,\sc{ii}$]\lambda3727$ & 715, 30 & 23.7, 0.8  & 0.0 \\
        $[$Ne\,\sc{v}$]\lambda3426$ & 731, 216 & 14.7, 1.0  & \cdots \\
        $[$Ne\,\sc{v}$]\lambda3426_{\rm core}$ & 476, 234 & 3.7, 3.1  & -5.4, {48} \\
        $[$Ne\,\sc{v}$]\lambda3426_{\rm wing}$ & 1747, 351 & 11.0, 3.2  & 204, {350} \\
        Mg\,\sc{ii}\,$\lambda2796_{\rm broad}$ &  2064, 257  &  54.2, 4.5  & 54, 57 \\
        \noalign{\smallskip}
        \multicolumn{1}{c}{$R_{4570}$$^{a}$}  &  \multicolumn{3}{c}{0.76$\pm$0.04} \\
        \noalign{\smallskip}
        \hline
    \end{tabular}
    \parbox[]{\columnwidth}{%
    {\bf Note.} 
    Column (2): The FWHM of each line, not corrected for instrumental broadening, in units of \kmps.
    Column (3): The flux of each line in units of $10^{-16}$\flux. 
    Column (4): The relative velocity of each line with respect to \oii\ in units of \kmps.\\
    $^{a}$ The profile and redshift are tied to \oiii\ core component. \\
    $^{b}$ The profile and redshift are tied to \hb\ broad component. \\
    $^{c}$ The relative strength of \feii, $R_{4570}\equiv$Fe\,{\scshape ii}\,$\lambda4570$/H$\beta_{\rm total}$, where Fe{\scshape\,ii}\,$\lambda4570$ is calculated by integrating the \feii\ multiplet from 4434\AA\, to 4684\AA.
    }
\end{table}

\subsection{Neil Gehrels \swift\ observatory}

In order to measure the X-ray spectrum, 
search for X-ray variability on long (years) and short (days) timescales, 
and to obtain the first simultaneous optical--UV--X-ray SED of 3C~286, 
we have obtained two observations with the Neil Gehrels \swift\ observatory \citep[\swift\ hereafter;][]{2004ApJ...611.1005G}
in 2020 (Table~\ref{tab:obs-log}; Target-Id 13644, PI: S. Komossa). 

\subsubsection{\swift\ XRT} 

The \swift\ X-ray telescope \citep[XRT;][]{2005SSRv..120..165B}
was operating in photon counting mode \citep[][]{2004SPIE.5165..217H}
with exposure times of 1--2\,ks each  (Table~\ref{tab:obs-log}).
During the two observations, the countrates of 3C~286 were low and constant. 
For the spectral analysis, we extracted the source photons within a circle of radius 47\arcsec, 
while the background was determined in an annulus with inner and outer radius of 60\arcsec\ and 100\arcsec\ respectively.
As 3C~286 is faint in X-rays, we combined the data sets of the two single observations in order to carry out the spectral fits.  
We created new ancillary response files (arfs) by adding the arfs of the single spectra weighted by their exposure time using the {\sc{ftool}} command {\it addarf}. 
The co-added X-ray spectra in the band $0.3-10\rm\,keV$ have only 38 counts in total.
Thus the spectra were re-binned to have 7 counts in each bin and then analyzed based on $C$-statistics \citep[][]{1979ApJ...228..939C} using the software package {\sc xspec} \citep[version 12.11.1;][]{1996ASPC..101...17A}. 
A single power law with Galactic absorption of $N^{\rm Gal}_{\rm H}=1.11\times10^{20}\rm\,cm^{-2}$ \citep[][]{2016A&A...594A.116H} fits the data well. 
The spectrum is flat with photon index $\Gamma_{\rm x}=1.8\pm0.5$ ($C/\rm dof=4.64/3$ using $C$-statistics).
The absorption-corrected flux is $F_{0.3-10\rm\,keV}=1.4^{+0.6}_{-0.5}\times10^{-13}$\,\flux, 
corresponding to a luminosity of $4.9\times10^{44}$\,\lum.

\begin{table}
\footnotesize
    \caption[]{The results of single power law model fitted to the X-ray spectra.
    }
    \label{tab:xray_spec}
    \centering                          
    \begin{tabular}{c l l l c}        
    \hline\hline                 
    \multicolumn{1}{c}{ } & 
    \multicolumn{1}{c}{$\Gamma_{\rm X}$} & 
    \multicolumn{1}{c}{$A_{1\rm\,keV}$$^{a}$} & 
    \multicolumn{1}{c}{$f_{0.3-10\rm\,keV}$$^{b}$} & 
    \multicolumn{1}{c}{$C$/dof} \\
    \hline                        
    \noalign{\smallskip}
    \swift & $1.8\pm0.5$ & $2.2^{+0.5}_{-0.6}$ & $1.4_{-0.5}^{+0.6}$ & 4.6/3 \\
    \noalign{\smallskip}
    \chandra & $2.1\pm0.2$ & $12.2^{+1.2}_{-1.1}$ & $6.4\pm0.6$ & 10.4/11 \\
    \chandra$^{c}$ & $2.2\pm0.2$ & $12.6^{+1.2}_{-1.1}$ & $7.4^{+0.7}_{-0.6}$ & 11.4/11 \\
    \noalign{\smallskip}
    \hline                                   
\end{tabular}
\parbox[]{\columnwidth}{%
    {\bf Note. }
    Galactic absorption with $N_{\rm H}^{\rm Gal}=1.11\times10^{20}\rm\,cm^{-2}$ is always included in the model during the fitting.\\
    $^{a}$ The normalization of the power law at the observed frame at $1\rm\,keV$ in units of $10^{-5}\rm\,photons\,keV^{-1}\,cm^{-2}\,s^{-1}$. \\
    $^{b}$ The Galactic absorption-corrected flux in units of $10^{-13}$\,\flux. \\
    $^{c}$ An extra absorption at redshift $z=0.692$ with column density of $N_{\rm H}=2\times10^{21}\rm\,cm^{-2}$ and 10\% solar abundances is added.
    }
\end{table}

\subsubsection{\swift\ UVOT}

Two observations of 3C~286 were obtained with the UV-optical telescope \citep[UVOT;][]{2005SSRv..120...95R}
in all six filters (Table~\ref{tab:obs-log})
in order to obtain SED information. 
For each observation, data in each filter were co-added using the task {\em{uvotimsum}}.
Source counts in all filters were then extracted in a circular region of radius 5\arcsec, while the background was selected in a nearby source-free region of radius 15\arcsec.
The background-corrected counts were then converted into 
fluxes based on the latest calibration as described in \citet{2008MNRAS.383..627P} and \citet{2010MNRAS.406.1687B}, using 
the task {\em{uvotsource}} and {\sc{caldb}} (version 20201026).
We checked the derived $b-v$ colours and find that they are out of the range over which the conversions used by {\em{uvotsource}} are applicable.
Thus we fit the UV/optical photometric data with a log-parabolic model, 
and calculate new conversion factors, 
as well as the Galactic extinction correction, 
by folding the best-fit log-parabolic model with the effective area of each filter following \citet{2010A&A...524A..43R}.
The final flux is obtained using the new conversion factors.

\subsection{\chandra\ data} 
We have also analyzed archival \chandra\ data 
in order to search for flux and/or spectral variability in
comparison with the \swift\ data. 
3C~286 was observed by \chandra\ ACIS-S for an exposure time of 2\,ks on February 26 2013 as a Guest Observer program (Table~\ref{tab:obs-log}; observation ID 15006, PI: J. Kuraszkiewicz).
The data are reduced using {\sc{ciao}} (version 4.12) and {\sc{caldb}} (version 4.9.2.1). 
The level 2 event file is created following the standard procedure.
Source detection performed using the {\em{celldetec}} task reveals the 
center for the X-ray source at $\rm R.A.=13^{\rm h}$31$^{\rm m}$08.31$^{\rm s}$, $\rm Dec.= +30^{\circ}$30$'$33.0$''$, 
with an offset of only 0.3\arcsec\ from the optical position of 3C~286.
No other X-ray source is detected near 3C~286 within $\sim$2\,\arcmin. 
We have searched for extended X-ray emission, but find that 95\% of the photons are within 2\,\arcsec, consistent with a point source. 

\begin{figure}
  \centering
  \includegraphics[width=0.9\columnwidth]{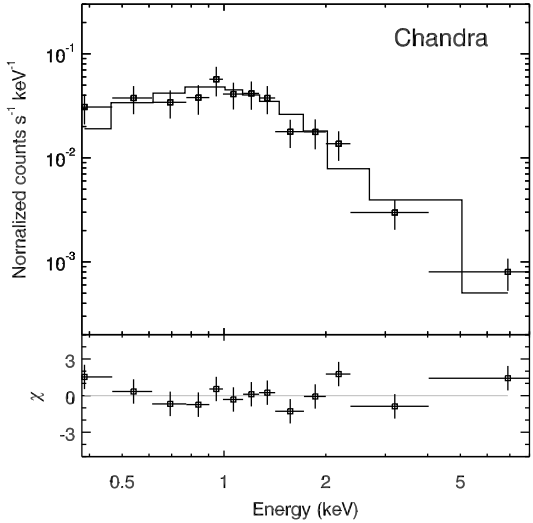}
  \caption{
  Power-law model fit to the \chandra\ ACIS-S spectrum of 3C~286 (upper panel), 
  and the residuals (lower panel) of the power-law fit to the spectrum. 
  }
  \label{fig:swift-residuals}
\end{figure}

For spectral analysis, we extract source photons from a circular 
region with radius of 6\arcsec. 
The background is determined in an annulus region with inner and outer radius of 10\arcsec\ and 30\arcsec, respectively. 
As there are only 130 net counts, 
the spectrum is grouped to have at least 10 counts per bin and
the $C$-statistics is adopted for minimization.
The spectrum is well fit by a single power law with absorption fixed at the
the Galactic value and a photon index $\Gamma_{\rm x} = 2.1\pm0.2$ 
($C/\rm dof=10.4/11$ using $C$-statistics).
No excess absorption is required (Figure~\ref{fig:swift-residuals}). 
The photon index agrees with the value measured with \swift\ within errors,
while the absorption-corrected flux $F_{0.3-10\rm\,keV}=(6.4\pm0.6)\times10^{-13}$\,\flux\ 
varied by a factor of $\sim4$ compared to \swift\ observation. 
Since a low-metallicity damped Ly$\alpha$ absorption system with $N_{\rm H}\sim2\times10^{21}\rm\,cm^{-2}$ was found at $z=0.692$ in front of 3C~286 \citep[e.g.,][]{1973ApJ...184L...7B, 1992ApJ...399L.121M},
we also test an additional intervening absorption by $N_{\rm H}=2\times10^{21}\rm\,cm^{-2}$ with 10\% solar metallicity at redshift of $z=0.692$ (see Section~\ref{sec:absorption}), 
but find a worse fit.
The parameters of models for both \swift\ and \chandra\ are listed in Table~\ref{tab:xray_spec}.

\subsection{Archival data} 
\label{sec:archivaldata}

\begin{figure*}
  \centering
  \includegraphics[width=0.8\textwidth]{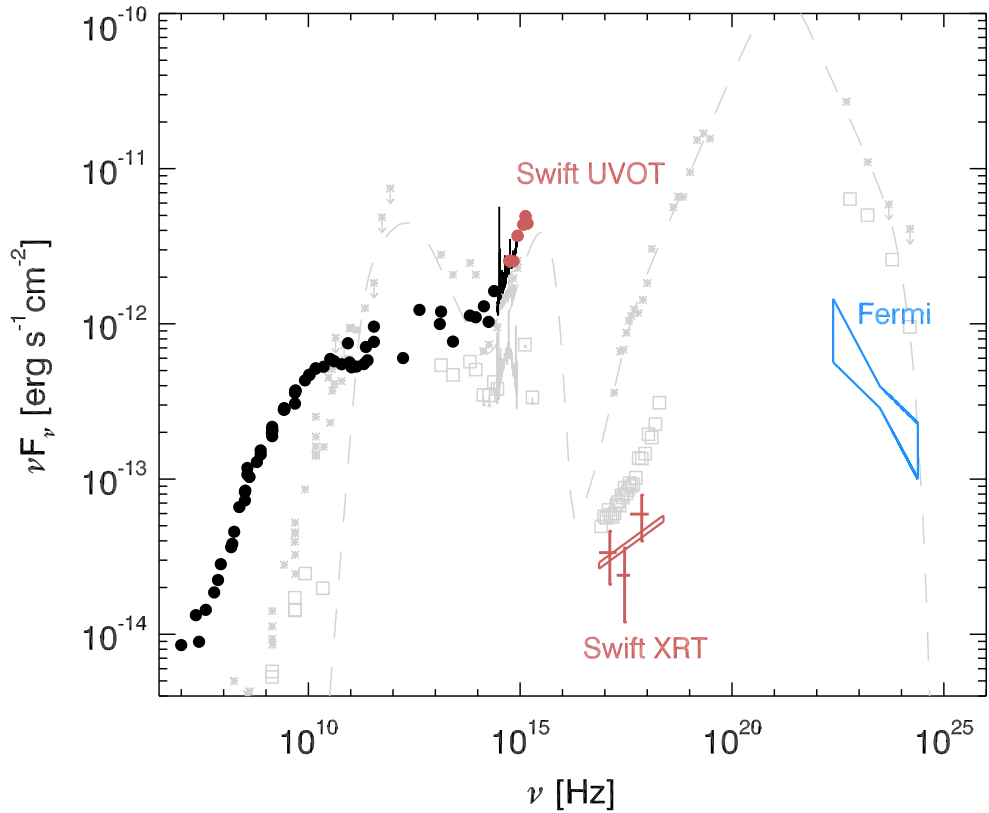}
  \caption{SED of 3C~286 based on our simultaneous \swift\ UVOT and XRT data (red) corrected for the Galactic absorption/extinction, along with non-simultaneous measurements from the \fermi\ Fourth Source Catalog (blue polygon),
  SDSS-BOSS spectroscopy (black curve) 
  and the literature (filled black circles, see Section~\ref{sec:archivaldata}). 
  The XRT data are re-binned for visual clarity and the red line represents the best-fit single power law.
  The SEDs of two $\gamma$-ray detected NLS1s at similarly high redshift, J0946+1017 (grey open squares) and J1222+0413 (grey asterisk), and the model for J1222+0413 (grey dashed line) are also plotted for comparison.
  }
              \label{fig:SED}
\end{figure*}

For an assessment of the IR variability and determination of the broad-band SED of 3C~286, 
first, we have inspected data from the {\em Single Exposure Source Table} of WISE/NeoWISE \citep[][]{2010AJ....140.1868W, 2011ApJ...731...53M} and find no evidence for significant variability
on both short and long time-scales.
We then used the 
average measurements from combined images reported in the AllWISE Source Catalog \citep[][]{2013wise.rept....1C}.
In addition, the infrared data from 2MASS \citep[][]{2006AJ....131.1163S} and data from {\it Spizter} \citep[][]{2007ApJ...660..117C} are included to build the broad-band SED.
We have also added 
the \fermi\ spectrum in $100\rm\,MeV$-$10\rm\,GeV$ based on the measurements from the \fermi\ Fourth Source Catalog \cite[][]{2020ApJS..247...33A}, 
the SDSS-BOSS UV-optical spectrum (Section~\ref{sec:sdss_spec})
and
the core radio fluxes taken from
\citet{1968Afz.....4..129A},
\citet{1980MNRAS.190..903L},
\citet{1981A&AS...45..367K},
\citet{1991ApJS...75.1011G},
\citet{1995ApJ...450..559B},
\citet{1996MNRAS.282..779W},
\citet{1997A&AS..122..235L},
\citet{2001A&A...372..719M},
\citet{2004ApJ...612..974C},
\citet{2004MNRAS.352..563C},
\citet{2007AJ....134.1245C},
\citet{2007A&A...470...83G},
\citet{2009A&A...502...61M},
\citet{2009ApJ...694..222C},
\citet{2010ApJ...714.1170M},
\citet{2010MNRAS.401.1240J},
\citet{2011ApJS..194...29R},
\citet{2011ApJS..192...15G}
and 
\citet{2014A&A...566A..59A}.
The SED of 3C~286 is shown in Figure~\ref{fig:SED}.
For comparison, we plot the SEDs of two $\gamma$-ray detected NLS1 galaxies at similar redshift, J0946+1017 \citep[$z=1.004$,][]{2019MNRAS.487L..40Y} and J1222+0413 \citep[$z=0.966$,][]{2015MNRAS.454L..16Y}. 
The model of J1222+0413's SED taken from \citet{2015MNRAS.454L..16Y} is also shown.
As can be seen from Figure~\ref{fig:SED}, 
compared to its optical/UV flux, 
3C~286 is relatively weak in X-rays.

\subsection{SMBH mass and Eddington ratio} 

We use three different methods of estimating BH mass. 
First, 
the SMBH mass $M_{\rm BH}$ can be estimated using the monochromatic continuum luminosity at 5100\,\AA, $L_{5100}$, and broad line width of \hb\ from the Gauss fit, 
which are proxies for the radius and the virial velocity of the BLR, respectively \citep[][]{2006ApJ...641..689V}.
Assuming that the emission of 3C~286 at 5100\,\AA\ is fully due to the disk emission,
we measure $L_{5100}=4.7\times10^{45}$\,\lum\ and obtain $M_{\rm BH}=2.3\times10^{8}\rm\,M_{\odot}$ 
\citep[Equation 5 in][]{2006ApJ...641..689V}.
Then, we obtained an Eddington ratio, defined as the ratio of bolometric luminosity to Eddington luminosity, 
of $\lambda_{\rm Edd}=1.6$ assuming a bolometric correction of $k=9.8$ \citep[][]{2004MNRAS.352.1390M}.
Alternatively, 
considering the SED in the optical may be partly contributed by jet, 
we estimate $L_{5100}=1.6\times10^{45}$\,\lum\ from the \hb\ luminosity based on the relation given in \citet{2006ApJS..166..128Z}, 
which leads to $M_{\rm BH}=1.3\times10^{8}\rm\,M_{\odot}$ and $\lambda_{\rm Edd}=1.0$.
Finally,  a third method was used, based on the finding   
that the width $\sigma$ of the \oiii\ core component
is a good surrogate for stellar velocity dispersion $\sigma_{*}$,
where $\sigma=\rm FWHM/2.35$ 
\citep[][]{2000ApJ...544L..91N,2007ApJ...667L..33K}.  
Employing the $M_{\rm BH}$-$\sigma_{*}$ relation \citep[][]{2005SSRv..116..523F}, and 
based on our decomposition of \oiii, we obtain $M_{\rm BH}=4.3\times10^{8}\rm\,M_{\odot}$. 
As above, we use $L_{5100}$ as upper limit, and the same bolometric
correction. Then, $\lambda_{\rm Edd}=0.9$.

All values of Eddington ratios from the 3 estimates above are consistent with the result $\lambda_{\rm Edd}=0.9$ obtained from the SED modelling in \citet{2020ApJ...899....2Z}, 
indicating that 3C~286 is accreting near the Eddington limit.

\section{Discussion}

\subsection{AGN type from optical spectral classification}

Because of its high redshift, the \hb\ regime of 3C 286 was not easily accessible in the past. 
A spectrum by 
\citet{1994ApJS...91..491G}
included \hb, but the line was too noisy for measurements. 
H$\alpha$ was detected by 
\citet{2003MNRAS.346.1009H},
but could not be resolved from neighbouring [N{\sc\,ii}] emission lines, with a total FWHM of $\sim$2600\,\kmps.
3C~286 was previously classified as a broad-line Seyfert 1.5 \citep[][]{2010A&A...518A..10V}
and recently suggested to be of NLS1-type based on the presence of \feii\ emission and an analysis of the \hb-[O{\sc\,iii}] complex of the SDSS-BOSS spectrum
\citep[e.g.,][]{2016PhDT.yao,2017FrASS...4....8B, 2020MNRAS.491...92L}.
Here, we present
for the first time a full analysis of the UV-optical spectrum of 3C~286 from SDSS-BOSS including all detected emission lines.
Given the unique properties of 3C~286, a reliable optical classification and an assessment of its variability is of great interest, and is essential in understanding the physical processes which govern its central engine. 
We discuss several classifications in turn: 

\subsubsection{Absorbed intermediate-type broad-line Seyfert 1 galaxy}\label{sec:absorption}

Given evidence for absorption along our line of sight,
the question arises, whether the inner BLR of 3C~286 is highly obscured, such that we therefore miss a large part of the high-velocity component in H$\beta$, or whether the whole spectrum is affected by intervening dusty absorbers weakening it enough that the broadest component of H$\beta$ becomes undetectable. 
Either would then imply, that the NLS1 classification is uncertain or even incorrect. 

First, there is no evidence for {\em intrinsic} absorption/extinction in 3C~286. Its UV spectrum is very blue, 
the X-ray spectrum only requires Galactic absorption, 
and the H{\sc\,i} 21\,cm absorption intrinsic to 3C~286 is exceptionally low with $N_{\rm H}<0.052\times10^{20}\rm\,cm^{-2}$ \citep[][]{2019ApJS..245....3G}. 

However, 3C~286 is well known for its damped Ly$\alpha$ system (DLA) from an {\em intervening} absorber at $z$=0.692 detected in UV spectra 
\citep[e.g.,][]{1992ApJ...399L.121M,2008Natur.455..638W}
and in HI 21 cm absorption 
\citep[e.g.,][]{1973ApJ...184L...7B,1976ApJ...208L..47W}
with a column density of 
$N_{\rm H}\sim2\times10^{21}$\,cm$^{-2}$
\citep[][]{1994ApJ...421..453C}.
The redshift and the very narrow Hydrogen absorption line width argues against an origin in 3C~286 itself. Detection of diffuse emission 2.5\arcsec\ from 3C~286 suggests that the intervening absorber is a low surface brightness galaxy 
\citep[][]{1994AJ....108.2046S}.
If the absorber of high column density was dusty, it would then significantly extinct the spectrum of 3C~286. 
However, measurements of multiple absorption lines from Hydrogen and different metals have shown, that both the metal abundances and also the dust to gas ratio of the absorber are very low. 
\citet{1992ApJ...399L.121M} report a metallicity of only 6\% solar and a dust-to-gas ratio of only 5\% of that of the Galactic disk. 
A low dust content of the absorber was also inferred from  the blue UV spectrum of 3C~286 
\citep[][]{1998A&A...333..841B}, 
and is consistent with larger sample studies of DLAs which find that these generally have low dust-to-gas ratios 
\citep[][]{1991ApJ...378....6P,1997ApJ...478..536P}.

We therefore conclude, that 3C~286 does not suffer strong extinction and that we measure the true H$\beta$ width. 
As a note in passing, 
the low metallicity of the intervening absorber likely explains, why our X-ray spectral fit is consistent with Galactic absorption.

\subsubsection{Starburst contribution} 

Given that \feii\ emission complexes are also detected in some strong starbursts 
and that [O{\sc\,ii}] is strong in the optical spectrum, 
the question arises if starburst emission could be responsible for the \feii\ complexes 
\citep[e.g.,][]{1994ApJ...427..174L}.
However, the profile of [O{\sc\,ii}] is similar to that of [O{\sc\,iii}]$_{\rm core}$ arguing for a NLR origin, 
the lack of any detectable absorbing material argues against significant (gas-rich) starburst activity, and the IR emission should be dominated the synchrotron emission of the jet, not starburst-heated dust \citep[see also][]{2020ApJ...899....2Z}.

\subsubsection{NLS1 classification and outflow component}

Taken at face value, line width measurements of \hb\ (Gauss or Lorentz) both imply a NLS1 classification of 3C~286 with FWHM(\hb, Gauss) = 2001\,\kmps\ and FWHM(\hb, Lorentz) = 1858\,\kmps. 
However, the width of \hb\ turns out to be very similar to broad [O{\sc\,iii}]. 
This is unusual and demands caution, as broad blue-shifted [O{\sc\,iii}] represents an outflow component, 
and the question is raised, whether \hb\ could be part of that same outflow and not representing the actual BLR, then just mimicking a NLS1 classification. 
However, other studies of outflows in NLS1 galaxies \citep[e.g.,][]{2008ApJ...680..926K, 2016A&A...591A..88B, 2018MNRAS.477.5115K}
have shown, that only a negligible fraction of \hb-emitting matter (typically below the detection limit) and of other low-ionization gas participates in the outflow, 
which is dominated by high-ionization transitions like [O{\sc\,iii}], [Ne{\sc\,iii}] and [Ne{\sc\,v}]. 
The fact that Mg{\sc\,ii} at rest has a similar FWHM as \hb\ then argues for the fact that both transitions represent emission from a bona-fide BLR, confirming the NLS1 classification. 

Finally, it has been suggested occasionally, that SMBH masses in NLS1 galaxies are higher then usually inferred, because of projection effects (a near face-on view onto a flattened BLR). 
While this effect was found not to play a major role in the majority of NLS1 galaxies 
(e.g., Section 7.2 of \citealt{2018rnls.confE..15K} for a review), 
it could still be relevant in individual objects. 
However, 
3C~286 is seen at a large angle of 48$^{\circ}$ \citep[][]{2017MNRAS.466..952A} rather than face-on, 
and therefore projection effects do not play a significant role in narrowing down the observed line widths.

We therefore conclude, that 3C~286 can be reliably classified as NLS1-type, taken all evidence from the full optical spectrum combined with multi-wavelength data and especially the tight constraints on intrinsic absorption.

\subsection{Radio-loudness}

Since no excess absorption/extinction is detected in 3C~286, the optical flux measurement is reliable. 
Further, the bulk radio emission of 3C~286 is very compact. 
Therefore, 
the radio-loudness index defined as $R=f_{5\rm\,GHz}/f_{4400\,\AA}$ in \citet[][]{1989AJ.....98.1195K}
can be reliably measured.
Using the SDSS $g$-magnitude of 17.33\,mag and the VLA $5\rm\,GHz$ flux of $7.5\rm\,Jy$ \citep[][]{1997A&AS..122..235L},
we find $\log R=4.4$ after $k$-correction with radio spectral index $\alpha_{\rm rad}=-0.61$ \citep[][]{2017MNRAS.466..952A} and optical spectral index $\alpha_{\rm opt}=-0.24$ from our SDSS spectrum analysis.
If excluding the extended radio emission in the radio-index estimate,
and instead using just the VLBI 4.9\,GHz flux of 1.55\,Jy of the brightest component of the core region on a scale of only $\sim50\rm\,pc$ \citep[][]{2017MNRAS.466..952A},
then the radio loudness would still be $\log R=3.7$.
We also calculated the radio loudness at 1.4\,GHz
(which was often used in work on samples of NLS1 galaxies) 
using the FIRST 1.4\,GHz flux of 15\,Jy \citep[][]{1995ApJ...450..559B}. 
This then gives $\log R_{1.4\rm\,GHz}=4.7$. 

Therefore, 3C~286 is the radio-loudest NLS1 galaxy known; with a radio-loudness index even high when compared to broad-line radio galaxies (which show values between $10^{2}$ to $10^{5}$
\citep[e.g.,][]{2007ApJ...658..815S}).

\subsection{Outflow component}

3C~286 shows additional blue-wing components in its high-ionization emission lines [O{\sc\,iii}], [Ne{\sc\,iii}] and [Ne{\sc\,v}] (Figure~\ref{fig:fblines}), 
implying that those lines arise in two different regions, a classical NLR and an outflow component. 
Blue wings are seen in a subset of other radio-quiet NLS1 \citep[e.g.,][]{2000ApJ...536L...5S,2004ApJ...606L..41G,2008ApJ...680..926K} 
and radio-loud NLS1 galaxies \citep[][]{2016A&A...591A..88B,2018MNRAS.477.5115K}, too.
Drivers of the outflow can be jet-cloud interactions or radiation-pressure driving, for instance \citep[][]{2018rnls.confE..15K}.
Given that the jet in 3C~286 is rather inactive with constant polarization over decades, 
and given the high Eddington ratio of 3C~286 ($L/L_{\rm Edd} \sim 1$), we tentatively favor a large-scale outflow over local jet-cloud interactions.

\subsection{Implications from emission-line ratios}

Individual emission-line ratios are good estimators of the physical conditions of the line-emitting gas, and the shape of the SED. 
In our case, we can use the line ratio [O{\sc\,ii}]/[O{\sc\,iii}] as an estimator of the ionization parameter $U$, 
and the ratio of [Ne{\sc\,iii}]/[Ne{\sc\,v}] as an indicator of the EUV SED (EUV excess or deficit), following \citet[][their Section~3.2 and Figure~2]{2006ApJ...639..710K}.
An intensity ratio $\log($[O{\sc\,ii}]/[O{\sc\,iii}]$)=-0.75$ implies an ionization parameter of the NLR of $\log U=-2.4$. 
In combination with the measured intensity ratio $\log($[Ne{\sc\,iii}]/[Ne{\sc\,v}]$) = 0.11$,  
this provides evidence for a soft excess 
in the EUV SED of 3C~286 (continuum ``c4'' in Figure~2b of \citealt{2006ApJ...639..710K}), 
consistent with the sharp rise we actually observe in the UV part of the SED of 3C~286 (Figure~\ref{fig:SED}). 
We note in passing, that these parameters are very similar to the radio-loud NLS1 galaxy SDSS~J172206.03+565451.6 \citep[][]{2006ApJ...639..710K}.

\subsection{SED} 

For the first time, we have measured with \swift\ a simultaneous optical--UV--X-ray SED of 3C~286 (Figure~\ref{fig:SED}).
The sharp rise in the optical-UV likely implies a strong
accretion-disk contribution in the EUV 
\citep[see also][]{2020ApJ...899....2Z}, 
as observed in other (radio-quiet) NLS1 galaxies, too. 
However, given the high redshift of the source, 
unfortunately the soft X-rays are redshifted out of the \swift\ band. Therefore, it can not be directly evaluated whether there is a soft X-ray excess which is commonly thought as a connection between the EUV emission from the accretion disk and the hard X-ray emission.
On the other hand, 
the faintness of the hard X-rays can be seen from the comparison of 3C~286 with two other $\gamma$-ray detected NLS1 galaxies at similar redshift in Figure~\ref{fig:SED}:
3C~286 has a higher flux in the optical-UV but a much lower flux in the hard X-ray band than J0946+1017 and J1222+0413.

We calculated the optical-to-X-ray spectral slope defined as $\alpha_{\rm ox}=-0.384\log(f_{2\rm\,keV}/f_{2500\,\AA})$
\citep[][]{1979ApJ...234L...9T} for 3C~286 using the rest-frame flux densities at 2\,keV and 2500\,\AA\ obtained from the \swift\ XRT and UVOT $w1$ band (effective wavelength $\sim$2500\,\AA) after $k$-correction.
The result, $\alpha_{\rm ox}=1.93$, indicates a steep optical-to-X-ray spectral slope \citep[][]{2010ApJS..187...64G}
and relatively faint hard X-rays ($>1\rm\,keV$) in comparison to the optical-UV emission.
This is also shown in comparison to two $\gamma$-ray (blazar) NLS1 galaxies, which we overplotted for comparison in Figure~\ref{fig:SED}.
The flatness of the X-ray spectrum itself implies the dominance of the jet emission in the hard X-ray band, 
likely composed of a mix of synchrotron and inverse Compton emission, as the spectral index measured with \chandra\ is steeper than the inverse Compton limit. 
Coronal emission from an accretion disk may contribute, too.

\subsection{Variability}

Since 3C~286 is an important calibrator source in the radio band with constant radio emission across many radio observations \citep[][]{1994A&A...284..331O, 2013ApJS..204...19P} on the one hand, 
but on the other hand hosts a 
NLS1 galaxy with a potentially highly variable accretion disk/corona system as seen in other NLS1s \citep[][]{2018rnls.confE..34G},
we have searched for long- and short-term variability in the X-ray regime. 
For this, we used archival {\it Einstein} data, 
our analysis of archival \chandra\ data, and our \swift\ XRT data (Table~\ref{tab:xray_var}).

\begin{table}
\footnotesize
    \caption[]{X-ray variability results. 0.5--4.5\,keV X-ray fluxes of 3C~286 measured with Einstein, Chandra and Swift. 
    }
    \label{tab:xray_var}
    \centering                          
    \begin{tabular}{l c c c c}        
    \hline\hline                 
    \multicolumn{1}{l}{date} & 
    \multicolumn{1}{c}{1980-06-30} & 
    \multicolumn{1}{c}{2013-02-26} & 
    \multicolumn{1}{c}{2020-08-16/21}  \\
    \hline                        
    \noalign{\smallskip}
    flux$^{a}$ & $5.3\times10^{-13}$ & $4.0^{+0.4}_{-0.3}\times10^{-13}$ &
    $8.5^{+2.3}_{-2.3}\times10^{-14}$ \\
    \noalign{\smallskip}
    \hline                                   
\end{tabular}
\parbox[]{\columnwidth}{%
    $^{a}$ corrected for Galactic absorption, in units of \flux. 
    }
\end{table}

3C~286 was observed by the {\it Einstein} observatory in 1980 and the flux in the 0.5--4.5\,keV band corrected for Galactic absorption was reported as $5.3\times10^{-13}$\,\flux\ \citep[][]{1983ApJ...268...60T}.
In the same energy range, the absorption-corrected \swift\ XRT and \chandra\ fluxes were 
$(8.5\pm2.3)\times10^{-14}$\,\flux\ and 
$4.0^{+0.4}_{-0.3}\times10^{-13}$\,\flux, respectively. 
Even considering the large uncertainties of the {\it Einstein} flux, 
these numbers imply that the observed X-rays varied by a factor of $\sim4$ or higher over the past decades.
As there is no evidence for absorption in the host galaxy, variability must be intrinsic and either from the accretion disk or jet. 
Given the evidence for $\gamma$-ray variability based on the \fermi\ data \citep[][]{2020ApJ...899....2Z},
the variable X-rays are then most likely related to the process which also causes the $\gamma$-ray variability, 
even though an origin in the accretion-disk corona cannot be excluded at present.

\subsection{Origin of the $\gamma$-ray emission and NLS1-character}\label{sec:gammaorig}

The majority of the $\gamma$-ray detected AGNs are blazars with relativistic jet directed to the observer \citep[][]{2020ApJS..247...33A}.
Their $\gamma$-ray emission is produced by relativistic particles, 
typically electrons, 
in the innermost jet upscattering seed photons either from the synchrotron radiation of the jet itself (synchrotron-self-Compton process)
or from the accretion disk, BLR and torus (external Compton process)
\citep[][]{2015A&ARv..24....2M}.
Unlike blazars, radio galaxies 
are ``misaligned'' AGNs viewed at larger angle to the line of sight, 
and they are interpreted as the parent population of blazars in the unification model of radio-loud AGNs \citep[][]{1995PASP..107..803U}. 
So, then the question is raised, how to produce the $\gamma$-ray emission in 3C~286. 
We address several scenarios.

First of all, it is important to note, that there is an intervening galaxy along our line of sight toward 3C~286 detected as DLA system (Section~\ref{sec:absorption}). 
We therefore wondered, whether that galaxy could instead be the counterpart to the \fermi-detected $\gamma$-ray emission. 
However, we can reject this scenario. 
First, the X-rays coincide with the optical position of 3C~286 within 0.3\arcsec\ and no second source is detected at 2.5\arcsec\ offset. 
Second, the intervening low surface brightness galaxy was found to have very low metal abundances, 
while any AGN/blazar is characterized by solar or even super-solar metallicity.

Next, we consider a scenario in which 3C~286 is a merger and harbors a binary SMBH; one of them accounting for the NLS1--CSS character and the other one being a blazar and producing the detected $\gamma$-rays. 
However, we do not find evidence for systematic emission-line shifts or even two systems of emission lines, 
and there is no evidence for excess absorption intrinsic to 3C~286 which argues against a gas-rich merger which harbors two active SMBHs.

Next, we ask if 3C~286 has an inner jet pointing at us, and responsible for strong beaming, but an outer jet bent to another direction 
\citep[][]{2015A&ARv..24....2M}. 
However, from the apparent proper-motion speed, \citet{2017MNRAS.466..952A} have found that the inner jet has an inclination angle to the line of sight of $\theta=48^{\circ}$ on a few tens of parsec scale, then bends at a distance of $\sim600$\,pc.
This shows, that the inner jet is still not pointing directly at us. 

Next, we note that 3C~286 is not the only CSS so far detected in $\gamma$-rays. 
A few other CSSs (young and evolving radio galaxies) have also been identified as $\gamma$-ray emitters in recent years though their number is very low  \citep[][]{2020ApJS..247...33A,2020ApJ...899....2Z}.
The scenario for their $\gamma$-ray production is still under debate.   
We recall that flat radio spectrum NLS1 galaxies are actually considered as CSSs viewed at small angles \citep[e.g.,][]{2001ApJ...558..578O, 2006MNRAS.370..245G,2006AJ....132..531K,2008ApJ...685..801Y,2016A&A...591A..98B}. 
Therefore, the mechanism discussed in the previous papers on $\gamma$-ray emission of CSS can be directly applied to 3C~286, too. 

For instance, it was suggested that the high energy emission up to GeV energies could be produced from the upscattered photons of an accretion disk and/or torus by relativistic electrons injected from hot spots into 
expanding lobes \citep[][]{2008ApJ...680..911S}.  Alternatively, the shock-accelerated particles in
the young lobes could initially yield bright bremsstrahlung emission in the GeV range and then fade out \citep[][]{2009MNRAS.395L..43K,2011MNRAS.412L..20K}.

In particular, authors suggested that,  
even though
faint, the $\gamma$-ray emission is still due to beaming of a jet at intermediate viewing angle \citep[][]{2011ApJ...740...29K}, 
which leads to a lower Doppler boosting, as indeed found by modelling the broad-band SED \citep[][]{2020ApJ...899....2Z}. 
This scenario is well possible, as long as the episodes of mild beaming do not dominate the total observed radio emission, which has been known to be so constant over some past decades. 
Another variant of this same scenario is, that individual jet components get occasionally deflected toward the observer as they encounter single dense blobs in the ISM along the path of the jet 
\citep[][]{2012A&A...539A..69B,2012ApJ...755..170B}.

Finally, it is also possible that external comptonization contributes to the $\gamma$-ray emission, and that it is variable because the NLS1 character of 3C~286 
(low black hole mass, high Eddington ratio) leads to a strong and variable external photon field. 

The last scenarios can be tested by long-term monitoring of 3C~286 with \fermi, 
along with quasi-simultaneous optical spectroscopy and SED measurements with \swift\ and in the VLBI radio regime.

\section{Summary and conclusions} 

We have carefully established the NLS1 character of 3C~286 based on its full UV-optical spectrum and multi-wavelength data, 
and based on a variety of measurements and considerations which go far beyond a simple FWHM measurement.  
Of particular importance for the source classification was evaluating complexity in the [O{\sc\,iii}] line implying a strong outflow component, not seen in 
Mg{\sc\,ii} however,
as well as the multiple lines of evidence for a lack of significant dust extinction towards 3C~286 and the lack of intrinsic absorption.

The simultaneous optical--UV--X-ray SED implies a strong accretion-disk component in the EUV, 
consistent with emission-line ratio diagnostics
([O{\sc\,ii}]/[O{\sc\,iii}] in combination with [Ne{\sc\,iii}]/[Ne{\sc\,v}]). 

With a SMBH mass of order $10^8\rm\,M_{\odot}$, 3C~286 is at the high-mass range of NLS1 galaxies, and at the low-mass end of radio-loud BLS1 galaxies. 
It is accreting near the Eddington limit, which may explain the strong outflow component we detect in all high-ionization emission lines.

3C~286 is the radio-loudest $\gamma$-ray emitting NLS1 galaxy identified so far, with a radio loudness of $\log R=4.4$, and one of the radio-loudest systems known. 

Our detection of high-amplitude X-ray variability (plausibly associated with those jet-component(s) which also emit the $\gamma$-rays) 
suggests caution when using 3C~286 as radio calibrator in highest-resolution radio VLBI observations.

While 3C 286 is not an EHT calibrator as it is too extended, it would become an interesting primary target in the future, if the flaring in X-rays and $\gamma$-rays re-occurs.

\section*{Acknowledgements}

We would like to thank the \swift\ team for carrying out our observations, and Jose L. {G{\'o}mez} for very useful discussions. 
Su Yao acknowledges support by an Alexander von Humboldt Foundation Fellowship. 
This work has made use of the data products from SDSS, \swift\ and \chandra.
This work has also made use of the NASA Astrophysics Data System Abstract Service (ADS), and the NASA/IPAC Extragalactic Database (NED) which is operated by the Jet Propulsion Laboratory, California Institute of Technology, under contract with the National Aeronautics and Space Administration.

Funding for the SDSS IV has been provided by the
Alfred P. Sloan Foundation, the U.S. 
Department of Energy Office of 
Science, and the Participating 
Institutions. 
SDSS-IV acknowledges support and 
resources from the Center for High 
Performance Computing  at the 
University of Utah. The SDSS 
website is www.sdss.org.
SDSS-IV is managed by the 
Astrophysical Research Consortium 
for the Participating Institutions 
of the SDSS Collaboration including 
the Brazilian Participation Group, 
the Carnegie Institution for Science, 
Carnegie Mellon University, Center for 
Astrophysics | Harvard \& 
Smithsonian, the Chilean Participation 
Group, the French Participation Group, 
Instituto de Astrof\'isica de 
Canarias, The Johns Hopkins 
University, Kavli Institute for the 
Physics and Mathematics of the 
Universe (IPMU) / University of 
Tokyo, the Korean Participation Group, 
Lawrence Berkeley National Laboratory, 
Leibniz Institut f\"ur Astrophysik 
Potsdam (AIP),  Max-Planck-Institut 
f\"ur Astronomie (MPIA Heidelberg), 
Max-Planck-Institut f\"ur 
Astrophysik (MPA Garching), 
Max-Planck-Institut f\"ur 
Extraterrestrische Physik (MPE), 
National Astronomical Observatories of 
China, New Mexico State University, 
New York University, University of 
Notre Dame, Observat\'ario 
Nacional / MCTI, The Ohio State 
University, Pennsylvania State 
University, Shanghai 
Astronomical Observatory, United 
Kingdom Participation Group, 
Universidad Nacional Aut\'onoma 
de M\'exico, University of Arizona, 
University of Colorado Boulder, 
University of Oxford, University of 
Portsmouth, University of Utah, 
University of Virginia, University 
of Washington, University of 
Wisconsin, Vanderbilt University, 
and Yale University.

\section*{Data Availability}

The data underlying this article are available in the SDSS archive at 
\url{https://dr16.sdss.org/optical/spectrum/search}, 
the astronomical archives of the HEASARC at
\url{https://heasarc.gsfc.nasa.gov/docs/archive.html},
the \swift\ archive at \url{https://swift.gsfc.nasa.gov/archive/} 
and the \chandra\ data archive at
\url{https://cda.harvard.edu/chaser/}, 
and can be accessed with the source coordinates or observation IDs.



\bibliographystyle{mnras}
\bibliography{3c286_ref} 








\bsp	
\label{lastpage}
\end{document}